\documentclass[twocolumn,pre,amssymb,floatfix]{revtex4}

\usepackage{times}
\usepackage{subfigure}
\usepackage[sort&compress]{natbib}
\usepackage{epsfig}
\usepackage{graphicx}

\usepackage{fancyhdr}
\usepackage{latexsym}
\usepackage{graphicx}
\usepackage{amsmath,amsmath,amsfonts}

\def\wintras{\omega}
\def\wintrab{\bar{\omega}}
\def\wintrainfty{\omega^{\infty}}

\def\wintrap{P}
\def\wintrapinfty{\wintrap^{\infty}}

\def\gtotb{\bar{g}_{\rm tot}}
\def\gtots{g_{\rm tot}}
\def\gtotp{g_{\rm tot}}
\def\gtotinfty{g_{\rm tot}^{\infty}}

\def\ginterb{\bar{g}_{\rm inter}}
\def\ginters{g_{\rm inter}}
\def\ginterinfty{g_{\rm inter}^{\infty}}

\def\zlattice{z_{\rm latt}}
\def\zc{z_{c}}
\def\yc{y_{c}}
\def\wc{w_{c}}
\def\ze{z}
\def\ye{y}
\def\we{w}
\def\zeinfty{\ze^{\infty}}
\def\yeinfty{\ye^{\infty}}

\def\Gauss{\rm id}

\def\dV{b}
\def\volfrac{\phi}
\def\As{A}
\def\Bs{B}
\def\gend{\delta g_{\rm end}}
\def\gendsum{d_{\rm end}}
\def\chie{\chi_{\rm e}}

\def\la{\langle}
\def\ra{\rangle}

\def\rv{{\bf r}}

\def\Rv{{\bf R}}

\def\cmon{c}

\graphicspath{{figs/}}

\begin{document}

\title{On the chain length dependence of local correlations in polymer melts \\
       and a perturbation theory of symmetric polymer blends}

\author{David C. Morse}
\affiliation{ 
   Department of Chemical Engineering \& Materials Science,
   University of Minnesota, 421 Washington Ave. S.E.,
   Minneapolis, MN 55455 }

\author{Jun Kyung Chung} 
\affiliation{ 
   Department of Physics,
   University of Minnesota, 116 Church St.,
   Minneapolis, MN 55455 }
\date{\today}

\begin{abstract}
The self-consistent field (SCF) approach to the thermodynamics of dense polymer 
liquids is based on the idea that short-range correlations in a polymer 
liquid are almost independent of how monomers are connected into polymers 
over large scales. Some limits of this idea are explored in the context 
of a perturbation theory for mixtures of two structurally identical polymer 
species, $\As$ and $\Bs$, in which the $\As\Bs$ pair interaction differs 
slightly from the $\As\As$ and $\Bs\Bs$ interaction, and this difference 
is controlled by a parameter $\alpha$. An expansion of the free energy to 
first order in $\alpha$ yields an excess free energy of mixing of the form 
$\alpha \ze(N)\volfrac_{\As}\volfrac_{\Bs}$, in both lattice and continuum 
models, where $\ze(N)$ is a measure of the number of inter-molecular near 
neighbors of each monomer in a one-component ($\alpha=0$) reference liquid 
with chains of length $N$. This quantity decreases slightly with increasing 
$N$ because the self-concentration of monomers from the same chain around 
each monomer is slightly higher for longer chains, creating a deeper 
inter-molecular correlation hole for longer chains. We present a theoretical 
analysis of the resulting $N$-dependence of local liquid structure, and 
predict that $\ze(N) = \zeinfty [1 + \beta \bar{N}^{-1/2}]$, where $\bar{N}$ 
is an invariant degree of polymerization, and $\beta = (6/\pi)^{3/2}$ is 
a universal coefficient. This and related predictions are confirmed by 
comparison to simulations of a continuum bead-spring model, as well as to 
previously published lattice Monte Carlo simulations. We propose a way to 
estimate the effective interaction parameter appropriate for comparisons 
of simulation data to SCF theory and to coarse-grained theories of 
corrections to SCF theory, which is based on an extrapolation of 
coefficients in this perturbation theory to the limit $N \rightarrow \infty$. 
We show that a renormalized one-loop theory contains a quantitatively 
correct description of the $N$-dependence of local liquid structure that 
we study here.
\end{abstract}
\maketitle             


\section{Introduction}
\label{sec:Intro}
In its most general form,\cite{Sanchez_89} the Flory-Huggins theory 
of polymer mixtures postulates that the free energy of mixing per 
monomer $\Delta f$ in a binary polymer mixture can be expressed as a 
sum of the form 
\begin{equation}
    \Delta f = kT 
    \sum_{i}\frac{\volfrac_i}{N_i}\ln \volfrac_{i} 
    + \Delta f_{\rm int}(\volfrac,T)
    \quad,
    \label{FloryHugginsDef}
\end{equation}
where $N_{i}$ is the degree of polymerization of species $i$, for 
$i=\As$ or $\Bs$,
$\volfrac_{\As} = \volfrac$ and $\volfrac_{\Bs} = 1 -\volfrac_{\As}$
are volume fractions, $kT$ is thermal energy, and 
$\Delta f_{\rm int}(\volfrac,T)$ is an interaction free energy 
per monomer. The original Flory-Huggins lattice model 
\cite{Flory_book_53} was more specific, insofar as it
predicted that $\Delta f_{\rm int}$ should be of the regular 
solution form $\Delta f_{\rm int} = C\volfrac_{\As}\volfrac_{\Bs}$, 
where $C$ is a temperature independent constant. In order to 
capture the variety of behaviors observed in real polymer 
mixtures, however, it has long been understood
\cite{Shibayama_Han_85,Stein_86,Han_88,Sanchez_89} 
that $\Delta f_{\rm int}$ must be allowed to exhibit an 
essentially arbitrary dependence on temperature and composition. 

This generalization of the Flory-Huggins theory is very flexible,
but not infinitely so: As long as $\Delta f_{\rm int}(\phi,T)$ 
is assumed to be independent of the chain lengths $N_{\As}$ and 
$N_{\Bs}$, the theory makes falsifiable predictions about how 
the phase diagrams for a homologous set of mixtures vary with 
$N_{\As}$ and $N_{\Bs}$, which are generally quite accurate. 
The assumption that $\Delta f_{\rm int}(\phi,T)$ is 
independent of chain length is the essential physical 
content of the theory.

The intuitive basis of this assumption is the idea that
$\Delta f_{\rm int}(\phi,T)$ represents a contribution to the
free energy that is sensitive only to the local structure of
a polymer liquid. It is also assumed that this contribution is 
separable from contributions that arise from the entropy of 
mixing in homogeneous mixtures and, in a generalization to
inhomogeneous liquids, from the distortion of polymer chains 
over large length scales.  The generalized Flory-Huggins theory 
described above is the homogeneous limit of a correspondingly 
general form of the self-consistent field theory (SCFT) of 
inhomogeneous polymer liquids. The key assumption in this 
SCFT \cite{Morse_Grzywacz_07} is that the interaction free 
energy density at any point in the fluid depends only upon 
the temperature and the average monomer concentration very 
near that point, independent of chain lengths, chain 
architecture, or compositions at distant points. 

The original Flory-Huggins lattice theory combined an
implicit assumption of locality (i.e., that local fluid 
structure is independent of chain length) with a much cruder 
random mixing approximation. The original theory introduced 
a random mixing approximation for the free energy of a 
lattice model in which monomers of types $i$ and $j$ on 
neighboring lattice sites interact with a potential energy 
$v_{ij}$. In the absence of vacancies, this yields
\begin{equation}
   \Delta f_{int} = 
   \alpha \zlattice \volfrac_{\As}\volfrac_{\Bs}
   \quad, \label{FloryHugginsRandomMixing}
\end{equation}
where $\alpha \equiv [ v_{\As\Bs} - (v_{\As\As}+v_{\Bs\Bs})/2 ]$, 
and where $\zlattice$ is the lattice coordination number, i.e., 
the number of lattice sites neighboring each site.

The random mixing approximation is known to substantially 
overestimate the actual energy of mixing for lattice models.
In simulations of a simple cubic lattice diluted with a 
modest density of vacancies, Sariban and Binder 
\cite{Binder_Sariban_87, Binder_Sariban_88} found that the 
energy of mixing was roughly half that predicted by an 
analogous approximation for a lattice with vacancies. In 
more recent lattice Monte Carlo simulations of diblock 
copolymer melts on a diluted fcc lattice, Matsen and 
coworkers \cite{Matsen_Vassiliev_03,Matsen_Vassiliev_06} 
also considered a lattice mean-field (i.e., random mixing) 
approximation for the order-disorder transition of symmetric 
diblocks, and found that it predicts a transition temperature 
more than twice that observed in their simulations, again 
indicating a large overestimation of the energy arising 
from $\As\Bs$ pair interactions. 

\subsection{Perturbation Theory}
For simple symmetric models, a much more accurate estimate 
for $\Delta f$ in the one-phase region can be obtained from 
thermodynamic perturbation theory. We focus here on a class 
of structurally symmetric models of $\As$-$\Bs$ polymer 
blends, on- or off-lattice, in which $\As$ and $\Bs$ chains 
are structurally identical, but in which the potential energy 
for $\As\Bs$ pair interactions is slightly different than 
that for $\As\As$ and $\Bs\Bs$ interactions.  Many of the 
coarse-grained models used in previous simulations of polymer 
blends fit this description, including some used in lattice 
Monte Carlo (MC) 
\cite{Binder_Sariban_87,Binder_Sariban_88,Binder_Deutsch_92b,
      Binder_Deutsch_93, Mueller_Binder_95}
off-lattice MC
\cite{dePablo_Gromov_98,Escobedo_dePablo_99}
and molecular dynamics (MD) 
\cite{Grest_Lacasse_96}
simulations.

To construct a perturbation theory, let $\alpha$ be a small 
parameter, with units of energy, that is proportional to the 
magnitude of the difference between $\As\Bs$ and $\As\As$ 
interactions.  Here, we consider an expansion of the free 
energy as function of $\alpha$ at constant temperature, in 
a liquid containing structurally identical $\As$ and $\Bs$ 
chains of equal length $N$. The limit $\alpha=0$ is thus a 
strongly correlated one-component liquid. At the critical 
composition $\volfrac=1/2$, phase separation occurs when 
$\alpha$ exceeds a critical value $\alpha_{c}$ that decreases 
as $1/N$ with increasing $N$.  An expansion of $\Delta f$ for 
$\alpha < \alpha_{c}$ to first order in $\alpha$ is thus 
expected to become increasingly accurate with increasing 
$N$.

In Sec. \ref{sec:Perturb} of this paper, we show that the
first order expansion for the free energy of mixing is given
by a function of the form
\begin{equation}
    \Delta f \simeq kT 
    \sum_{i}\frac{\volfrac_i}{N}\ln \volfrac_{i} 
    + \alpha \ze(N) \volfrac_{\As}\volfrac_{\Bs}
    \quad.
    \label{DeltaFPerturb}
\end{equation}
for both lattice and continuum models. In this expansion, 
the coefficient $\ze(N)$ is an "effective coordination 
number" whose value is sensitive to local correlations in 
the one-component reference state.  In a lattice model with 
nearest-neighbor interactions, $\ze(N)$ is found to be equal 
to the average number of inter-molecular nearest neighbors, 
i.e., the number of lattice sites neighboring a test monomer
that are occupied by monomers from a different chain than the 
one containing the test monomer. An analogous definition is
given for $\ze(N)$ in a continuum model. 

The main source of error in the random mixing model is the 
neglect of the consequences of the so-called correlation hole. 
The immediate environment of any monomer in a dense polymer 
liquid is crowded with other monomers from the same chain. In 
a nearly incompressible liquid, this causes a compensating 
depression in the number of neighboring monomers from other 
chains, leading to a correlation hole in the intermolecular 
distribution function, and thus to a decrease in the 
intermolecular interaction energy. The accuracy of a 
perturbation theory for the free energy of a system of
long chains is a result of the fact that the structure of 
this correlation hole at all $\alpha < \alpha_{c}$ is very 
similar to that in the $\alpha=0$ reference liquid, because 
of the smallness of $\alpha_{c}$.

Several authors have previously proposed approximations that 
are either equivalent to first order perturbation theory, or 
very closely related. 
M\"{u}ller and Binder \cite{Mueller_Binder_95} proposed a 
"modified Flory-Huggins" approximation for the free energy of 
mixing $\Delta f$ of a simple lattice model that is completely
equivalent to the first-order perturbation theory of Eq.
(\ref{DeltaFPerturb}).
In discussions of the results of continuum bead-spring 
simulations, both Grest and coworkers \cite{Grest_Lacasse_96} 
and Escobedo and de Pablo \cite{Escobedo_dePablo_99} have 
proposed approximations for structurally symmetric continuum 
models that are either equivalent \cite{Grest_Lacasse_96} or 
nearly equivalent \cite{Escobedo_dePablo_99} to the 
perturbation theory discussed here. None of these authors
explicitly described their proposed approximations as a
form of perturbation theory. 

M\"{u}ller and Binder \cite{Mueller_Binder_95} and
Escobedo and de Pablo \cite{Escobedo_dePablo_99} have both 
shown that these perturbative approximations provide extremely 
accurate predictions for the critical temperature $T_{c}$ 
observed in simulations of symmetric binary blends. In 
both of these studies, the authors showed that critical 
point found from finite-size scaling analysis of simulation
results is very well approximated by the expression
\begin{equation}
    \frac{\alpha\ze(N)N}{kT_{c}(N)}
    \simeq 2
\end{equation}
that is obtained by using the first-order expansion 
of Eq.  (\ref{DeltaFPerturb}) to predict $T_{c}$.
A somewhat stronger statement can be made about the 
lattice MC data of Binder, Mueller, and coworkers
\cite{Binder_Deutsch_93,Mueller_Binder_95,Mueller_95}:
Their results for the dimensionless parameter 
$\alpha \ze(N)N/kT_{c}(N)$ deviate systematically from
$2$ for finite chains, but appear to extrapolate in the
limit $N \rightarrow \infty$ to an asymptote that is 
equal to $2$ to within statistical uncertainties. 
This convergence was not emphasized by Binder
and coworkers, but appears to be a natural consequence 
of the identification of modified Flory-Huggins theory 
as first order perturbation theory, and the fact that 
deviations from any low order perturbation theory for 
$T_{c}$ must decrease with decreasing $\alpha_{c}$, and 
thus with increasing $N$.

\subsection{Chain Length Dependence of Local Correlations}
An important difference between the generalized Flory-Huggins 
theory of Eq. (\ref{FloryHugginsDef}) and perturbation theory
of Eq. (\ref{DeltaFPerturb}), which are superficially very similar, 
is that the coefficient $\ze(N)$ in Eq.(\ref{DeltaFPerturb}) 
actually does depend slightly upon the chain length $N$.  
M\"{u}ller and Binder \cite{Mueller_Binder_95} obtained values 
for $\ze(N)$ in bond fluctuation model simulations of 
one-component melts for several values of $N$, ranging from 
$N=$20,\ldots,160.  They found that $\ze(N)$ approaches a 
finite asymptote $\ze^{\infty}$ as $N \rightarrow \infty$, 
and that deviations of $\ze(N)$ from $\ze^{\infty}$ are 
well approximated by a function of the form
\begin{equation}
   \ze(N) = \zeinfty [ 1 + \beta \bar{N}^{-1/2} ]
   \quad. \label{zc_of_N}
\end{equation}
where $\bar{N} \equiv Nb^{6}/v^{2}$, $b$ is the statistical 
segment length, $v$ is the volume per monomer, and $\beta$ is 
a constant that M\"{u}ller and Binder determined empirically.

In Sec. \ref{sec:LengthDependence} of this paper, we analyze 
the chain length dependence of the short distance behavior of 
the intermolecular radial distribution function, and of 
related quantities such as $\ze(N)$. Our analysis starts from 
the (verifiable) assumption that, in a nearly incompressible 
liquid, the total radial distribution function, including both 
intra- and inter-molecular contributions, changes extremely 
little with changes in chain length $N$. The intramolecular 
correlation function does, however, change slightly with $N$, 
due to changes in the number of chemically distant monomers 
from the same chain that contribute to the self-concentration 
around any test monomer. This causes a systematic decrease
in the depth of the inter-molecular correlation function 
with decreasing $N$, and thus an increase in $\ze(N)$, as
as chemically distant intramolecular neighbors are simply
replaced by intermolecular neighbors in the immediate 
environment of any test monomer. The relevant concentration
of chemically distant intramolecular neighbors of any test 
monomer is calculated using a simple random walk model. 

We predict that, while $\zeinfty$ is a model dependent quantity, 
the coefficient $\beta$ in Eq. (\ref{zc_of_N}) has a universal 
value
\begin{equation}
   \beta = \left ( \frac{6}{\pi} \right )^{3/2} = 2.64
   \label{beta_value}
\end{equation}
for any structurally symmetric model. 

In Secs. \ref{sec:LatticeSimulations} and 
\ref{sec:ContinuumSimulations} we test our assumptions by  
comparing the resulting predictions to simulation results. In 
Sec. \ref{sec:LatticeSimulations}, we verify the accuracy of
Eq. (\ref{beta_value}) by comparing it to the lattice MC
results of M\"{u}ller and Binder. 
In \ref{sec:ContinuumSimulations}, we present a more extensive 
comparison to the results of our own off-lattice simulations,
which allow us to more directly test the assumptions 
underlying our analysis. 

\subsection{Comparing Blend Simulations to SCFT}
Simulations of dense polymer liquids are providing 
increasingly precise tests of the assumptions underlying SCF 
theory.  Lattice Monte Carlo and continuum simulations of 
simple coarse-grained models have been used to quantify slight 
deviations from the random walk model for polymer 
statistics in melts, 
and deviations from the RPA description of composition 
fluctuations in both polymer blends 
\cite{Binder_Sariban_87,Binder_Deutsch_93,Mueller_Binder_95,Grest_Lacasse_96}
and block copolymer melts
\cite{Binder_Fried_91a,Binder_Fried_91b,Pakula_97,Grest_Murat_99,Matsen_Vassiliev_03}.

In order to compare either simulation or experimental data 
to SCF predictions for multicomponent systems, however, one 
must somehow choose values for the SCF interaction free 
energy $\Delta f_{\rm int}(\phi,T)$ and/or the effective 
interaction parameter 
\begin{equation}
   \chi_{e} = \frac{-1}{2kT}
   \frac{\partial \Delta f_{\rm int}(\phi,T)}{\partial\phi^{2}}
   \label{chie_def}
\end{equation}
relevant to the RPA analysis of long-wavelength scattering. 
\cite{Sanchez_89}
[Note that, throughout this paper, $f$ and $\Delta f$ denote 
free energies per monomer, though we use the same symbols elsewhere 
to denote free energies per volume. \cite{Morse_Grzywacz_07}.]
Because the relationship between $\Delta f_{\rm int}(\phi,T)$ 
and the underlying microscopic parameters is never known 
{\it a priori}, the temperature and (sometimes) composition
dependence of $\Delta f_{\rm int}$ or $\chi_{e}$ have thus
far been determined by fitting RPA predictions to the 
available measurements of composition fluctuations, in the
analysis of either experiment or simulations.  
The uncertainty introduced by this fitting procedure becomes 
a potentially serious problem, however, when one's goal is 
to precisely quantify small deviations from RPA predictions, 
which is necessary in order to test theories that predict
corrections to SCFT, such as the one loop theory.  For this 
purpose, it is would be very useful to have independent way 
of unambiguously defining and accurately calculating 
$\Delta f_{\rm int}$ for a simulation model, using the
microscopic information that is available in a simulation.
Here, we propose a way of doing this for symmetric models,
which is based on an extrapolation of perturbation theory
to the infinite chain limit. 

SCFT has long been believed to be, in some sense, exact in 
the limit of infinitely long chains. The basis for this 
belief is, in part, the predictions of a variety of 
closely related one-loop theories of fluctuation effects. 
The one-loop theory 
\cite{deLaCruz_Edwards_88,Holyst_Vilgis_93,Holyst_Vilgis_94,
      Fredrickson_Liu_94,Wang_02,Morse_Grzywacz_07}
is a coarse-grained theory, that, when properly interpreted
\cite{Wang_02, Morse_Grzywacz_07}, predicts small corrections 
to the free energy of an underlying SCF theory. 
The relative magnitude of the fluctuation correction to the 
free energy is found 
\cite{Holyst_Vilgis_93,Holyst_Vilgis_94,Wang_02,Morse_Grzywacz_07}
to decrease as $N^{-1/2}$ with increasing chain length $N$,
implying that SCFT becomes exact as $N \rightarrow \infty$.
This idea makes sense, however, only if it is understood that 
the definition of SCF theory used in the construction of the 
renormalized one-loop theory \cite{Wang_02, Morse_Grzywacz_07}
is one in which $\Delta f_{int}$ is defined by a process of 
extrapolation to the infinite chain limit: That is, it makes 
sense to say that the true free energy $f$ convergences to 
some form of SCF free energy as $N \rightarrow \infty$ only 
if the relevant SCF is understood to be one whose parameters 
reflect the liquid structure of a hypothetical system of 
infinitely long chains. 

We propose here that a useful approximation for this 
asymptotic SCF free energy may be obtained for simple symmetric 
models by considering perturbative expansions of both 
$\Delta f_{int}$ and the true free energy $f$ in powers 
of $\alpha$. The idea is to identify the coefficients in 
the expansion of $\Delta f_{int}$ by considering the 
limiting behavior as $N \rightarrow \infty$ of the 
corresponding coefficients in the expansion of the true 
free energy for blends of finite chains. 

Specifically, we identify the first order contribution to 
$\Delta f_{\rm int}$ as the $N \rightarrow \infty$ limit 
of the first order contribution to Eq. (\ref{DeltaFPerturb}) 
for $\Delta f(\phi,N)$. This yields an SCF energy
\begin{equation}
  \Delta f_{int} \simeq
  \alpha \ze^{\infty} \volfrac_{\As}\volfrac_{\Bs}
  + {\cal O}(\alpha^{2})
  \label{DeltaFint}
\end{equation}
to first order in $\alpha$.
This first order approximation yields the same dependence 
on composition as the original Flory-Huggins theory, but 
reflects the local correlations present in a hypothetical 
reference system of infinitely long chains. 

In Sec. \ref{sec:CoarseGrained}, we discuss the relationship 
between the ${\cal O}(N^{-1/2})$ contributions to $\ze(N)$ 
in the first order perturbation theory for finite $N$ and 
the corrections to SCF theory predicted by a renormalized 
one-loop theory. To test the compatibility of one-loop 
predictions with our analysis of the perturbation theory, 
we consider the predictions of the one-loop theory for the 
value of the derivative $\partial \Delta f/\partial \alpha$ 
in a symmetric blend with $\alpha=0$. We find that the one-loop 
prediction for this quantity is actually {\it identical} to 
that given by Eqs.  (\ref{DeltaFPerturb}), (\ref{zc_of_N}), 
and Eq. (\ref{beta_value}), with the same value for the 
numerical coefficient $\beta$ of the ${\cal O}(N^{-1/2})$ 
correction. This implies that the renormalized one-loop 
theory correctly describes the chain length dependence of 
the correlation hole that is discussed here.

\section{Models and Notation}
\label{sec:Models}
In what follows, we consider a general class of structurally 
symmetric binary polymer blends, uisng a similar language for
lattice and continuum models.  We consider a system containing 
a total of $M$ structurally identical chains, each containing
$N$ monomers, in which $\volfrac_{\As}M$ are of type $\As$ and 
$\volfrac_{\Bs}M$ are of type $\Bs$.  Let $\alpha$ be a small 
parameter that controls the magnitude of the difference between 
$\As\Bs$ and $\As\As$ interactions. The state $\alpha=0$ is 
thus a ideal mixture of $M$ physically indistinguishable chains,
in which a fraction $\volfrac_{\As}$ can be chosen to be $\As$
chains at random.

\subsection{Lattice Models}
We consider a class of lattice models in 
which double occupancy of lattice sites is forbidden, and in 
which monomers on neighboring sites of types $i$ and $j$ interact 
with a pair potential $v_{ij}(\alpha)$ of the form 
\begin{equation}
   v_{ij}(\alpha) = u + \alpha \dV_{ij} \quad.
\end{equation}
Here, $u$ is the interaction between all neighboring monomers, 
$\alpha$ is a small parameter, and $\dV_{ij}$ is a symmetric 
matrix of dimensionless coefficients, with 
$\dV_{\As\Bs} = \dV_{\Bs\As}$.
To maintain the symmetry between the two species, we require 
that $\dV_{\As\As} = \dV_{\Bs\Bs}$.  
The value of the parameter $u$ is relevant if and only if the 
system contains vacancies, because changes in $u$ can then 
effect correlations in the one-component reference liquid.  

Binder, Deutsch, and M\"{u}ller
\cite{Binder_Deutsch_93,Mueller_Binder_95,Mueller_95}
have simulated one variant of the bond fluctuation model in 
which (in our notation) $\dV_{\As\As} = 0$ and $\dV_{\As\Bs}=1$ 
and another in which $\dV_{\As\Bs} = 1/2$ and $\dV_{\As\As}=-1/2$.

\subsection{Continuum Models}
We also consider a class of structurally symmetric continuum 
models.  Consider a system of $M$ chains of length $N$ in a 
volume $V$, giving an overall monomer concentration $c = MN/V$ 
or an average monomer volume $v = 1/c$.  The total potential 
energy is the sum of intramolecular bonding potentials, which 
are assumed to be the same for $\As$ and $\Bs$ chains, plus a 
sum of nonbonded pair potentials.  The pair potential for 
monomers of type $i$ and $j$ separated by a distance $r$ is 
assumed to be of the form
\begin{equation}
   v_{ij}(r) = u(r) + \alpha \dV_{ij}(r)
   \quad,
\end{equation}
with $\dV_{\As\As}(r) = \dV_{\Bs\Bs}(r)$, where $\alpha$ has units
of energy.

In the continum model of Grest {\it et al} \cite{Grest_Lacasse_96},
$v_{ij}(r)$ is taken to be a purely repulsive shifted Lennard-Jones 
(LJ) interaction, with the same LJ diameter and cutoff distances for 
all pairs, but with an interaction energy that is slightly larger 
for $\As\Bs$ pairs than for $\As\As$ or $\Bs\Bs$ pairs. In this 
model, $\dV_{\As\As}(r) = 0$, and $u(r)$ and $\dV_{\As\Bs}(r)$ have 
the same functional form, differing only by a prefactor. A very 
similar model is used in our own continuum simulations, which are 
presented in Sec. \ref{sec:ContinuumSimulations}

\subsection{Correlation Functions}
To discuss a perturbation theory for continuum models, it is useful 
to introduce some notation for inter- and intra-molecular correlations 
in the the one-component reference liquid, with $\alpha=0$.  Let 
$\ginters(\rv,s,N)$ denote the intermolecular radial distribution 
function (RDF) for a test monomer with monomer index $s$ in a
reference liquid containing chains of length $N$, defined such that 
$\ginters(\rv,s,N) \rightarrow  1$ as $\rv \rightarrow \infty$. The 
product $c\ginters(\rv,s,N)$ is thus the probability density 
(probability per volume) of finding any monomer from a different 
chain separated by a vector $\rv$ from such a test monomer. Let 
$\wintras(\rv,s,N)$ be an intramolecular correlation function
in this reference liquid, defined as the probability density 
for finding any other monomer from the same chain separated 
by a vector $\rv$ from a test monomer with monomer index $s$. 
Let $\gtots(\rv,s,N)$ be the total RDF for a test monomer with 
a specific monomer index $s$, defined so that
\begin{equation}
   \cmon \gtots(\rv,s,N) = \wintras(\rv,s,N) + \cmon \ginters(\rv,s,N) 
   \quad,
\end{equation}
and so that $\gtots(\rv,s,N) \rightarrow 1$ as 
$\rv \rightarrow \infty$. 
Let $\ginterb(\rv,N)$, $\gtotb(\rv,N)$, and $\wintrab(\rv,N)$ 
denote the averages of $\ginters(\rv,s,N)$, $\gtots(\rv,s,N)$, 
and $\wintras(\rv,s,N)$, respectively, with respect to $s$.

\section{Perturbation Theory}
\label{sec:Perturb}
We consider a first-order perturbation theory for the free energy per 
monomer
\begin{equation}
   f \equiv \frac{-kT}{MN} \ln Z
\end{equation}
for either type of model. Here, 
\begin{equation}
 Z = \int D[\Rv] e^{-H/kT}
\end{equation}
is the partition function,
$H$ is the total potential energy and $\int D[\Rv]$ denotes an integral 
or (in a lattice model) sum over distinguishable microstates.  Differentiation 
of $f$ with respect to $\alpha$ yields
\begin{equation}
  \frac{\partial f}{\partial \alpha} = \theta
\end{equation}
where 
\begin{equation}
  \theta(\volfrac,\alpha) \equiv \frac{1}{MN}
  \left \langle \frac{\partial H}{\partial \alpha} \right \rangle
  \quad.
\end{equation}
To first order in $\alpha$, the free energy per monomer in a system
with a composition $\volfrac \equiv \volfrac_{\As}$, is thus given by
\begin{equation}
    f(\volfrac, \alpha) \simeq f_{0} +
    \frac{kT}{N}\sum_{i} \volfrac_{i}\ln \volfrac_{i} 
    + \alpha \theta(\volfrac,0)
\end{equation}
Here, $f_{0}$ is the free energy per monomer of a corresponding 
one-component reference state, with $\alpha=0$, when all of the 
chains are treated as indistinguishable in the calculation of
entropy. The ideal free energy of mixing term accounts for the 
combinatorial entropy associated with the random labelling of 
chains as $\As$ or $\Bs$.  The quantity $\theta(\volfrac,\alpha=0)$ 
is evaluated in the resulting ideal mixture.

In the simple case of a lattice model with $v_{\As\As}=u$ and 
$v_{\As\Bs}= u+\alpha$, $\theta$ is simply equal to the total
number of $\As\Bs$ neighbor pairs in the system, divided by 
the total number $MN$ of monomers. 

In a continuum model with $\dV_{\As\As}=\dV_{\Bs\Bs}=0$, $\theta$ 
is given by the sum of values of $\dV_{\As\Bs}(r)$ for all 
interacting $\As\Bs$ monomer pairs, divided by $MN$.

The composition dependence of $\theta$ within the ideal mixture 
can be determined by simple combinatorical arguments. To show 
this, we consider lattice and continuum models separately. 

\subsection{Lattice Models}
Let $\zc(N)$ be the average number of sites neighboring each monomer 
that are occupied by monomers from a different chain, evaluated in the 
reference state $\alpha=0$ (i.e., the average number of inter-molecular 
neighbors per monomer). Let $\wc(N)$ be the average number of neighboring
sites that are occupied by monomers from the same chain (the average 
number of intra-molecular neighbors). Let 
\begin{equation}
   \yc(N) = \zc(N) + \wc(N)
\end{equation} 
be the average total number of occupied neighbors. In the absence of 
vacancies, $\yc(N)$ must equal the latice coordination number.

In the reference state, the labelling of chains as type $\As$ or $\Bs$ 
is random.  The probability that any given pair of inter-molecular
neighbors will belong to different chains labelled with types $i$ 
and $j$, respectively, is thus simply $\volfrac_{i}\volfrac_{j}$. 
Similarly, the probability that any two intramolecular neighbors 
will belong to a chain of type $i$ is simply $\volfrac_{i}$. Thus, 
for $\alpha=0$,
\begin{equation}
  \theta = 
  \frac{1}{2} \we(N) \sum_{i} \volfrac_{i} \dV_{ii} +
  \frac{1}{2} \ze(N) \sum_{ij} \volfrac_{i} \volfrac_{j}\dV_{ij} 
  \quad,
\end{equation}
The prefactors of $1/2$ correct for the double counting of pairs.  
A bit of rearrangement yields
\begin{equation}
   \theta = \frac{1}{2}\ye(N) 
          + \ze(N)\volfrac_{\As}\volfrac_{\Bs}
   \quad, \label{theta_of_volfrac}
\end{equation}
where we have defined
\begin{eqnarray}
   \ye(N) & \equiv & \yc(N) \dV_{AA} \nonumber \\
   \ze(N) & \equiv & \zc(N) ( \dV_{AB} -\dV_{AA} ) 
   \quad.
\end{eqnarray}
The free energy per monomer is thus given to first order in $\alpha$ 
by
\begin{equation}
   f \simeq f_{0} + \frac{1}{2} \alpha \ye(N) + \Delta f
   \quad, \label{f_perturb}
\end{equation}
where $\Delta f$ is given by Eq. (\ref{DeltaFPerturb}), with the
definition of $\ze(N)$ discussed above. The resulting first order
expansion of $\Delta f$ is identical to the modified Flory-Huggins 
approximation of M\"{u}ller and Binder.\cite{Mueller_Binder_95}

\subsection{Continuum Models}
In a continuum model, for any value of $\alpha$,
\begin{equation}
   \theta = \frac{v}{2}\sum_{ij}
   \int d\rv \; \la c_{i}(\rv) c_{j}(0) \ra \dV_{ij}(\rv)
   \quad, \label{theta_continuum}
\end{equation}
where $c_{i}(\rv)$ is the instantaneous concentration of monomers 
of type $i$ and $v = V/MN$ is the volume per monomer.  In an ideal
mixture, with $\alpha=0$, random labelling of a fraction 
$\volfrac_{\As}$ of the chains as $\As$ and the remainder as $\Bs$ 
yields
\begin{equation}
    \la c_{i}(\rv) c_{j}(0) \ra = 
    \delta_{ij} c \wintrab(\rv,N) \volfrac_{i}
    + c^{2} \ginterb(\rv,N) \volfrac_{i}\volfrac_{j} \quad.
\end{equation}
Using expression in Eq. (\ref{theta_continuum}) yields an expression
for $\theta(\volfrac,0)$ of the same form as that given in Eq. 
(\ref{theta_of_volfrac}), and a free energy of the form given in 
Eqs.  (\ref{f_perturb}) and (\ref{DeltaFPerturb}), in which the 
quantities $\ye(N)$ and $\ze(N)$ are given for a continuum model 
by the integrals
\begin{eqnarray}
    \ye(N) & \equiv & \cmon
    \int d\rv \; \gtotb(\rv) \dV_{AA}(\rv)
    \nonumber \\
    \ze(N) & \equiv & \cmon
    \int d\rv \; \ginterb(\rv) \Delta\dV(\rv)
    \quad,
\end{eqnarray}
where
\begin{equation}
   \Delta\dV(\rv) \equiv \dV_{\As\Bs}(\rv) - \dV_{\As\As}(\rv) 
   \quad.
\end{equation}
In what follows, we will also consider the analogous quantities
\begin{eqnarray}
    \ye(s,N) & \equiv & \cmon
    \int d\rv \; \gtots(\rv,s,N) \dV_{AA}(\rv)
    \nonumber \\
    \ze(s,N) & \equiv & \cmon
    \int d\rv \; \ginters(\rv,s,N) \Delta\dV(\rv)
    \quad,
\end{eqnarray}
for a test monomer with a known monomer index $s$, with 
$1 \leq s \leq N$.

\section{Dependence of Local Liquid Structure on Chain Length}
\label{sec:LengthDependence}
In this section, we consider how properties of a one-component melt
that are sensitive to short range correlations, such as $\ze(N)$, 
depend upon overall chain length $N$. Our reasoning applies equally 
well to lattice and continuum models, but we will hereafter adopt a 
notation appropriate to a continuum model. To proceed, we first 
consider how the intramolecular correlation function $\wintras(\rv,s,N)$ 
depends upon $s$ and $N$, and then consider how this translates 
into a corresponding $s$- and $N$-dependence of $\ginters(\rv,s,N)$.

\subsection{Intra-molecular distribution}
Each monomer in the one-component reference liquid is surrounded 
by a concentration $\wintras(\rv,s,N)$ of other monomers from the 
same chain, in addition to a concentration $c\ginters(\rv,s,N)$ of
monomers from other chains. Let
\begin{equation}
    \wintras(\rv,s,N) \equiv \sum_{s'} \wintrap(\rv,s',s,N) 
\end{equation}
where $\wintrap(\rv,s',s, N)$ is probability density of finding 
a specific pair of monomers from the same chain, with monomer 
indices $s'$ and $s$, separated by a vector $\rv$.
The distribution $\wintras(\rv,s,N)$ is dominated, for small values 
of $\rv$, by contributions from monomers for which $|s'-s|$ is
small.  As a result, for monomers that are far from either chain 
end (i.e., far from $s=1$ and $s=N$), $\wintras(\rv,s,N)$ depends 
only weakly on chain length $N$ and index $s$. 

In the limit $N \rightarrow \infty$, $\wintrap(\rv,s,s',N)$ approaches 
a function 
\begin{equation}
    \wintrapinfty(\rv,|s-s'|) \equiv \lim_{N \rightarrow \infty}
    \wintrap(\rv,s,s',N)
\end{equation}
that depends only on $|s-s'|$, and $\wintras(\rv,s,N)$ approaches
a function
\begin{equation}
    \wintrainfty(\rv) \equiv 
    \lim_{N \rightarrow \infty} \wintras(\rv,s=N/2,N)
\end{equation}
that is independent of $s$, for $s$ not too close to either chain 
end.

\begin{figure}[tb]\center
\includegraphics[width=0.45\textwidth,height=!]{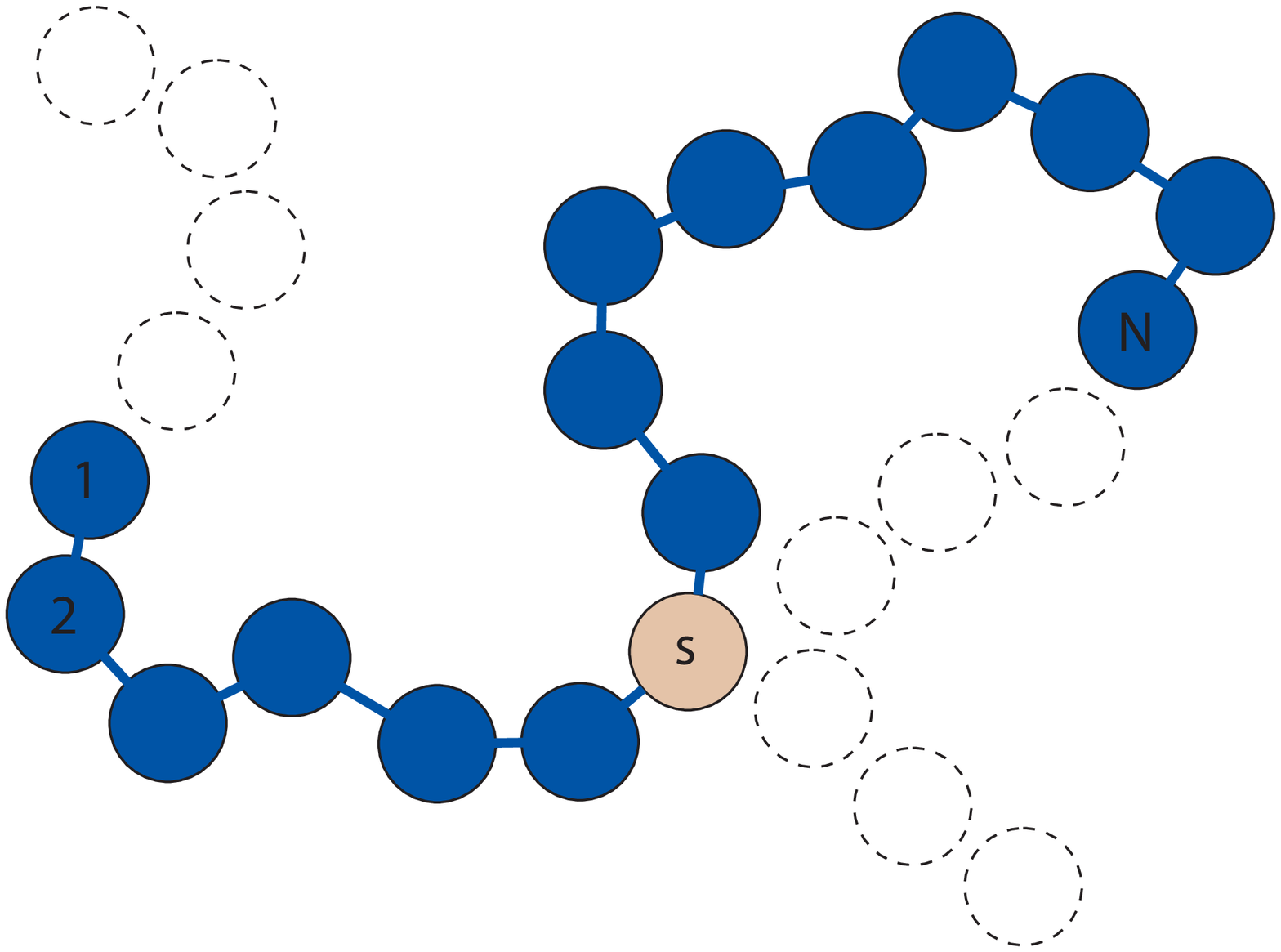}
\caption[Truncated chain]
{Schematic view of Eq. (\ref{wintras_sum_missing}) for the 
difference $\delta\wintras(\rv,s,N)$ between the intramolecular 
function $\wintrainfty(\rv)$ for an infinite chain and the 
corresponding correlation function
$\wintras(\rv,s,N)$ for monomer $s$ on a chain of length $N$. 
This difference is attributed to the contributions to 
$\wintrainfty(\rv)$ of the monomers $s \leq 0$ and $s > N$ 
that are "missing" from the finite chain. The concentration of 
these missing monomers near the test monomer (i.e., near $\rv=0$) 
can be estimated using a random walk model, if $s$ is not too
near either chain end.  As the chain length is decreased from 
$\infty$, chemically distant monomers from the same chain are 
simply replaced by monomers from other chains, while leaving 
the total RDF $\gtots(\rv,s,N)$ almost unchanged.}
\label{fig:chain}
\end{figure}

For finite chains, $\wintras(\rv,s,N)$ contains a small correction to 
$\wintrainfty(\rv)$, which depends upon both $s$ and $N$. Let
\begin{equation}
   \delta\wintras(\rv,s,N) \equiv \wintrainfty(\rv) - \wintras(\rv,s,N)
   \quad, \label{domega_def}
\end{equation}
where $\wintrainfty(\rv)$ and $\wintras(\rv,s,N)$ are evaluated at the 
same temperature and concentration.
To estimate $\delta\wintras(\rv,s,N)$, we assume that infinite chains 
and long finite chains have very similar conformational statistics. 
We thus approximate the difference $\delta \wintras(\rv,s,N)$ for 
monomers that are not too close to either chain end by the 
contributions to $\wintrainfty(\rv)$ from monomers with $s' \leq 0$ 
and $s' > N$. In this approximation,
\begin{equation}
 \delta \wintras(\rv, s,N) \simeq
       \sum_{s'=-\infty}^{0}  \wintrapinfty({\bf r},\Delta s)
    +  \sum_{s'=N+1}^{\infty} \wintrapinfty({\bf r},\Delta s)
 \quad, \label{wintras_sum_missing}
\end{equation}
where $\Delta s \equiv |s'-s|$. The content of this equation is 
shown schematically in Fig. \ref{fig:chain}.

For values of $\Delta s \gg 1$, we may further approximate 
$\wintrapinfty({\bf r}, \Delta s)$ by a Gaussian distribution
\begin{equation}
  \wintrap_{\Gauss}({\bf r}, \Delta s) = 
  \left ( \frac{3}{2\pi \Delta s b^{2}} \right )^{3/2}
  \exp \left ( -\frac{3 r^{2} }{2 \Delta s b^{2} } \right )
  \quad,
\end{equation}
for a continuous random walk, with $\Delta s = |s-s'|$. 
The use of random walk statistics is justified by the fact that, 
for $s \gg 1$ and $|N-s| \gg 1$, $|s'-s| \gg 1$ for every term 
in the relevant sums. In the same limit, we may also approximate 
sums over $s'$ by integrals to obtain an analytic approximation 
for $\delta\wintras_{\Gauss}(\rv,s,N)$.  Here and hereafter, a 
subscript $\Gauss$ (for ``ideal") is used to indicate approximations 
obtained using this idealized continuous random-walk chain model. 

The effective coordination number $\ze(s,N)$ is sensitive only 
to the distribution of monomers that lie within the range of the 
pair potential from a test monomer. To characterize how the 
self-concentration $\wintras(\rv,s,N)$ within this small region 
depends upon $s$ and $N$, we consider the $s$- and $N$-dependence 
of the distribution $\delta \wintras_{\Gauss}(\rv=0,s,N)$ for a 
random walk, evaluated at the the position ${\bf r}=0$ of the 
test monomer. The random-walk model yields a deviation 
\begin{eqnarray}
  \delta \wintras_{\Gauss}(0,s,N) & = &
  \left ( \frac{3}{2\pi b^{2}} \right )^{3/2}
  \left [ \int\limits_{s}^{\infty}\frac{d\Delta}{\Delta^{3/2}}
        +  \int\limits_{N-s}^{\infty}\frac{d\Delta}{\Delta^{3/2}}
  \right ] \label{domega0sNint}
   \nonumber \\
   & = &
   \left ( \frac{3}{2\pi} \right )^{3/2} 
   \frac{2}{b^{3}} 
   \left [ \frac{1}{\sqrt{s}} + \frac{1}{\sqrt{N-s}} \right ]
   \quad, \label{domega0sN}
\end{eqnarray}
where $\Delta \equiv |s-s'|$. The random walk approximation 
clearly breaks down for $s$ near the chain ends, as expected 
on physical grounds, since Eq. (\ref{domega0sN}) predicts a 
$1/\sqrt{s}$ divergence at either chain end.  

The free energy will be shown to depend upon on the average 
value
\begin{equation}
   \delta \wintrab(\rv,N) \equiv 
   \frac{1}{N}\sum_{s} \delta \wintras(\rv,s,N) 
   \quad.
\end{equation}
Using the above random walk model for $\delta\wintras(\rv=0,s,N)$ 
yields
\begin{equation}
   \delta \wintrab_{\Gauss}(\rv=0,N) =  \frac{1}{v}
   \left ( \frac{6}{\pi} \right )^{3/2}\frac{1}{\bar{N}^{1/2}}
\end{equation}
in which $\bar{N} \equiv N b^{6}/v^{2}$. The quantity
$v \delta\wintrab_{\Gauss}(\rv=0,N) = (6/\pi)^{3/2}\bar{N}^{-1/2}$ 
is the corresponding volume fraction of the "missing" monomers in 
the vicinity of a randomly chosen test monomer. 

Note that the integral with respect to $s$ required to calculate 
$\wintras_{\Gauss}(\rv=0,N)$ converges, despite the divergence of 
Eq. (\ref{domega0sNint}) for the integrand at both chain ends.
This reflects the fact that the average with respect to $s$ is 
dominated not by the contributions of a few monomers near the 
chain ends, but by that of many interior monomers.  As a result, 
our use of a random walk model is sufficiently accurate to correctly 
calculate the prefactor of the dominant ${\cal O}(N^{-1/2})$ to 
$\delta\wintras(\rv=0,s,N)$. Further corrections that arise from 
the breakdown of the random walk model near both chain ends are 
expected to yield subdominant contributions of ${\cal O}(N^{-1})$.  

\subsection{Inter-molecular distribution}
We now consider how the $N$-dependence of the intramolecular 
correlation function is reflected in intermolecular correlations.

\subsubsection{A simple argument for $\ze(N)$}
Each monomer in a dense melt is surrounded by an environment 
that contains both monomers from other chains and chemically
distant monomers from the same chain. Near any test monomer,
the local concentration of chemically distant monomers from 
the same chain is lower than that found for an infinite chain
by a concentration $\delta\omega(0,s,N)$, which is calculated
above. We assume that, in an incompressible liquid, this 
decrease in the self-concentration of a finite chain with
decreasing $N$ must be almost exactly compensated by a 
corresponding increase in the concentration of inter-molecular
near neighbors. We assume furthermore that, in a structurally
symmetric model with $\alpha=0$, the contribution of chemically
distant monomers on the same chain to the overall radial
distribution function $\gtots(\rv,s,N)$ within the small
region near a test monomer that lies within the range of
the potential is the same, per monomer, as the contribution
of monomers from other chains. Only intermolecular near
neighbors, however, contribute to the quantity $\ze(N)$. 
We thus expect a decrease in the local self-concentration 
by a volume fraction $v\delta\omega(0,s,N)$ is to increase 
the effective coordination number by a fractional amount
$v\delta\omega(0,s,N)$, due to a direct replacement of
chemically distant intramolecular neighbors by intermolecular
neighbors. This yields a predicted coordination number
\begin{equation}
   \ze(N) = \zeinfty [ 1 + v\delta\wintrab(0,s,N) ]
\end{equation}
or
\begin{equation}
   \ze(N) \simeq \zeinfty
   \left [ 1 + 
         \left ( \frac{6}{\pi}\right )^{3/2} 
         \frac{1}{\bar{N}^{1/2}} 
   \right ] \quad,
   \label{zeN_final}
\end{equation}
where $\zeinfty$ is the limiting value of $\ze(N)$ in the
limit $N \rightarrow \infty$.

\subsubsection{A more detailed view}
The same conclusion may be obtained from a set of more explicit 
assumptions about intra- and inter-molecular correlation functions 
in a dense liquid.  Let $\gtotinfty(\rv)$ and $\ginterinfty(\rv)$ 
denote the limits of the $\gtots$ and $\ginters$, respectively, in 
the limit $N \rightarrow \infty$.  The value of $\ze(N)$ depends 
only upon the behavior of the intermolecular distribution 
$\ginters(\rv,s,N)$ for small  $\rv$, for which $|\rv|$ is 
less than the range of the pair potential.  We make the 
following assumptions about how the small-$|\rv|$ behavior 
of $\gtotinfty(\rv,s,N)$ and $\ginters(\rv,s,N)$ depends 
upon chain length, in a dense melt in which the polymer coil 
size $\sqrt{N}b$ is much larger than the range of the pair 
potential:

(1) In an almost incompressible liquid, we assume that the 
total RDF $\gtots(\rv,s,N)$ is almost independent of both 
$s$ and $N$, except for monomers very close to one of the 
chain ends. In this problem, this is what is meant when
we say that the melt is effectively "incompressible". If 
true, this implies that 
\begin{equation}
   \gtots(\rv,s,N) \simeq \gtotinfty(\rv)
   \quad, \label{gtot_assume}
\end{equation}
for all monomers except a few near the chain ends.  
For this to be true for all $N$, however, the decrease in 
the intramolecular self-concentration $\wintras(\rv,s,N)$ 
with decreasing $N$ must be exactly compensated by an
increase in $\cmon \ginters(\rv,s,N)$. This implies
\begin{equation}
   \cmon \ginters(\rv,s,N) \simeq 
   \cmon \ginterinfty(\rv) + \delta\wintras(\rv,s,N)
   \quad, \label{ginter_assume}
\end{equation}
where $\delta\wintras(\rv,s,N)$ is defined by Eq. (\ref{domega_def}).

Corrections to assumption (\ref{gtot_assume}) can arise, even 
for values of $s$ that are far from either chain end, from 
contributions to $\gtotinfty(\rv,s,N)$ due to correlations 
between an interior monomer test monomer and an end monomer. The 
resulting end-effect corrections (discussed in more detail in 
the subsection) are of order $1/N$, and are found to be negligible 
compared to the $1/\sqrt{N}$ corrections that are obtained from 
the $N$ dependence of $\delta\wintras(\rv,s,N)$. 

(2) The spatial distribution around a test monomer of chemically 
distant monomers from the same chain is assumed to closely mimic 
the local distribution $\ginters$ of monomers from other chains, 
which we can approximate by the RDF $\ginterinfty(r)$ for a system 
of infinite chains. We thus assume that the distribution 
$\delta\wintras(\rv,s,N)$ of the chemically distant "mising" monomers 
(those that must be removed from an infinite chain to form the finite 
chain of interest) is of the form
\begin{equation}
   \delta\wintras(\rv,s,N) \propto \ginterinfty(\rv)
   \label{domega_propto_g}
\end{equation}
for large $N$, values of $s$ far from either chain end, and values 
$|\rv|$ less than the range of the potential.

(3) The constant of proportionality in Eq. (\ref{domega_propto_g}) 
depends on an overall concentration of ``missing" monomers over a
region larger than the range of the potential. We assume that an
average concentration of missing monomers near a test monomer can 
be obtained by using the random-walk model for the return 
probability $\delta\wintras(\rv=0, s, N)$. More precisely, we assume 
that 
\begin{equation}
   \delta\wintras(\rv,s,N) \simeq 
   \delta\wintras_{\Gauss}(0,s,N) \ginterinfty(\rv)
   \label{domega_assume}
\end{equation}
under the same conditions that Eq. (\ref{domega_propto_g}) applies.

By combining assumptions (\ref{ginter_assume}) and (\ref{domega_assume}),
we find that
\begin{equation}
   \ginters(\rv,s,N) \simeq \ginterinfty(\rv) 
   \left [ \; 1 + v  \delta\wintras_{\Gauss}(0,s,N) 
   \; \right ]
   \label{ginter_final}
\end{equation}
Using this approximation yields
\begin{equation}
   \ze(s,N) \simeq \zeinfty
   \left [ 1 + 
         v \delta \wintras_{\Gauss}(0,s,N)
   \right ] 
    \quad,
\end{equation}
where $\wintras_{\Gauss}(0,s,N)$ is given by Eq. (\ref{domega0sN}),
where 
\begin{equation}
    \zeinfty \equiv \cmon \int d\rv \; 
    \ginterinfty(\rv) \Delta\dV(\rv)
    \quad.
\end{equation}
Averaging with respect to $N$ yields Eq. (\ref{zeN_final}). 

\subsection{End Effects}
In addition to the dominant ${\cal O}(\bar{N}^{-1/2})$ corrections 
to $\ze(N)$ predicted above, we also expect to find subdominant 
${\cal O}(1/N)$ corrections to both $\ye(N)$ and $\ze(N)$ that
arise from true end effects, due to the perturbation of the liquid 
structure near chain ends. 

First, consider the contribution of end-effects to the total correlation 
function $\gtots(\rv,s)$, and to the corresponding integral $\ye(s,N)$.
Let $\gtotp(\rv,s,s',N)$ be a distribution function for pairs of 
monomers with specified monomer indices $s$ and $s'$, defined so that
$(c/N)\gtotp(\rv,s,s',N)$ is the probability per unit volume of finding 
any monomer with index $s'$ separated by $\rv$ from a test monomer with
index $s$, and so that
\begin{equation}
  \gtots(\rv,s) = \frac{1}{N}\sum_{s'}\gtotp(\rv,s,s',N)
\end{equation}
We assume that the deviation of $\gtotp(\rv,s,s',N)$ from $\gtotinfty(\rv)$ 
is dominated by pairs of values where either $s$ or $s'$ (but not both)
are near one of the chains, and the other is somewhere in the interior of 
the chain. Reflecting this assumption, we assume that $\gtotp(\rv,s,s',N)$
can be approximated by a function of the form
\begin{equation}
  \gtotp(\rv,s,s',N) = \gtotinfty(\rv) 
                     + \gend(\rv,s) + \gend(\rv,s')    
\end{equation}
where $\gend(\rv,s)$ is a deviation that is large only for $s$ near $1$ 
or $N$, and vanishes for interior monomers. This functional form assumes 
that the correction $\gend(\rv,s)$ that arises from $s$ near one of the 
chain ends is independent of $s'$ when $s'$ is an interior monomer, and
similarly for the correction $\gend(\rv,s')$. This approximation captures 
the dominant ${\cal O}(N^{-1})$ corrections to $\gtotb(\rv)$, but ignores 
smaller ${\cal O}(N^{-2})$ corrections arising from contributions in which
both $s$ and $s'$ are near chain ends. Within this ${\cal O}(N^{-1})$
approximation, evaluating the sum with respect to $s'$ yields
\begin{equation}
   \gtots(\rv,s) = \gtotinfty(\rv) + \gend(\rv,s) 
                + \frac{1}{N}\gendsum(\rv)
\end{equation}
where
\begin{equation}
   \gendsum(\rv) \equiv \sum_{s'}\gend(\rv,s')
   \quad.
\end{equation}
Within the above approximation, we thus predict that, for interior 
monomers, $\gtots(\rv,s)$ should deviate from $\gtotinfty(\rv)$ by 
an amount that is proportional to $1/N$ but independent of $s$. 

This approximation for $\gtots(\rv,s)$ also implies that we expect
to find 
\begin{equation}
   \ye(s,N) \simeq \yeinfty + \frac{\delta}{N}
   \quad, \label{yc_predict}
\end{equation}
for interior monomers. The $1/N$ correction to $\ye(s,N)$ for 
interior monomers is the result of occasional close contact 
between an interior test monomers and end monomers, and is 
assumed to be independent of the monomer index $s$ of the 
interior test monomer. In addition, we expect to see a much
larger ${\cal O}(1)$ deviation from $\yeinfty$ for the last 
few monomers at either chain end.  

Similar reasoning suggests that the quantity $\ze(s,N)$ 
for $s$ far from either chain end should also exhibit an 
${\cal O}(1/N)$ contribution, due to close contacts between 
the interior test monomer and end monomers of other chains, 
in addition to the ${\cal O}(N^{-1/2})$ correction described 
above. The same reasoning suggests that this ${\cal O}(1/N)$ 
correction for interior monomers should be independent of $s$,
implying a functional form
\begin{equation}
   \ze(s,N) \simeq \zeinfty
   \left [ \; 1 + v  \delta\wintras_{\Gauss}(0,s,N) \; 
   \right ] + \frac{\gamma}{N}
   \label{zesN_with_end}
\end{equation}
in which $\zeinfty$ and $\gamma$ are material parameters.

\section{Comparison to Lattice Simulations}
\label{sec:LatticeSimulations}

\begin{figure}[tb]
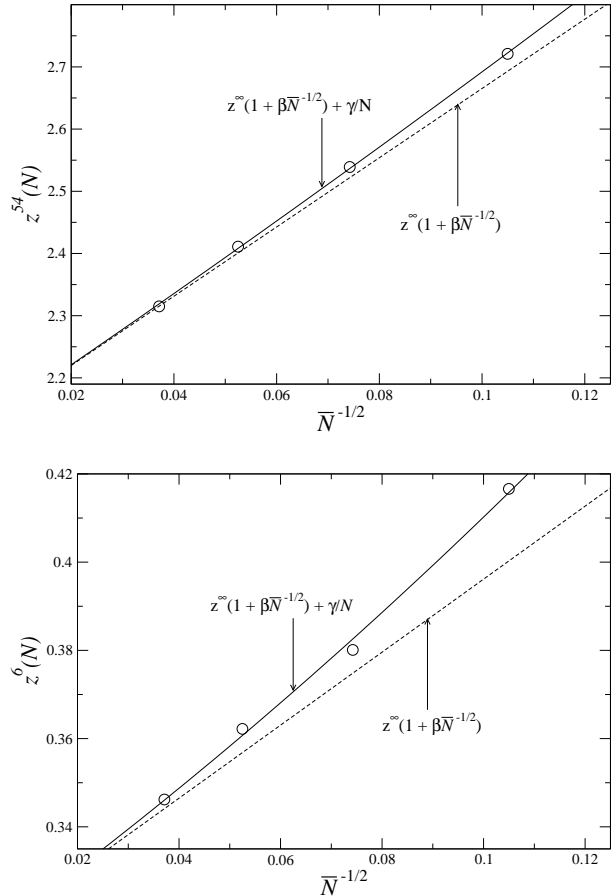
\center
\includegraphics[width=0.45\textwidth,height=!]{muller54.eps}

\vspace{12pt}

\includegraphics[width=0.45\textwidth,height=!]{muller6.eps}
\caption[$\ze(s,N)$]
{Lattice MC results of M\"{u}ller and Binder \cite{Mueller_Binder_95}
for $\ze(N)$ (symbols) vs. $\bar{N}^{-1/2}$ for two different 
variants of the bond fluctuation model. The top panel shows results
for a model in which each site interacts with monomers located on
any of 54 nearby sites, while the bottom shows results for a model 
in which each site can interact with only 6 neighboring sites. In
each panel, the solid line is a best fit to Eq. (\ref{zcAll_predict}), 
using the predicted value of $\beta=(6/\pi)^{3/2}$. This fit yields 
$\zeinfty=2.109$ and $\gamma'=0.588$ for the model with 54 neighbors 
and $\zeinfty=0.3134$ and $\gamma'=0.310$ for the model with 6 
neighbors. Dashed lines show the estimated asymptote 
$\zeinfty[ 1 + \beta \bar{N}^{-1/2}]$, with the same value for 
$\zeinfty$, in order to show the predicted asymptotic slope.}
\label{fig:muller}
\end{figure}

Fig. \ref{fig:muller} shows a comparison of theoretical 
predictions to the lattice Monte Carlo results of M\"{u}ller 
and Binder \cite{Mueller_Binder_95} for $\ze(N)$ for two 
different variants of the bond 
fluctuation model. In both variants of the model, a monomer
is taken to occupy 8 sites within a cubic lattice, from 
which other monomers are excluded. The volume fraction of
occupied sites is 50\%. The top panel of Fig. \ref{fig:muller} 
shows results for $\ze(N)$ for chains of length 
$N=$20, \ldots, 160 for a model in which each monomer interacts 
with monomers that are located at any of 54 neighbors. The 
bottom panel shows results for a model in which each monomer 
interacts only with monomers at the 6 nearest possible 
positions in a cubic lattice. We have compared both sets of 
data to a prediction
\begin{equation}
   \ze(N) \simeq \zeinfty
   \left [ 1 + \beta \bar{N}^{-1/2} \right ] + \gamma'/N
   \quad,
   \label{zcAll_predict}
\end{equation}
in which $\beta$ is given by Eq. (\ref{beta_value}), but in 
which $\zeinfty$ and $\gamma'$ are treated as adjustable 
parameters. The $\gamma'/N$ term is included to account both 
for the $1/N$ contribution to Eq. (\ref{zesN_with_end}) for 
interior monomers, and for contributions to the average over 
$s$ arising from ${\cal O}(1)$ deviations from $\zeinfty$ 
for monomers near either chain end. 

In the absence of a prediction for the coefficient $\beta$, 
M\"{u}ller and Binder fit each of these data sets to a function 
$\ze(N) = \zeinfty [ 1 + \beta_{fit} \bar{N}^{-1/2} ]$, while 
treating both $\zeinfty$ and $\beta_{fit}$ as adjustable 
parameters.  This yields best fit parameters 
$\beta_{fit} = 2.846$ for $\ze^{54}$ and $\beta_{fit} = 3.330$
for $\ze^{6}$ slightly higher than the predicted value of 2.64. 
The quality of the fit is approximately the same with either 
functional form. 

Our predictions of a univeral value for the asymptotic slope 
in these plots is consistent with this data. Inclusion of the 
$1/N$ end correction is necessary to adequately fit this data
for modest values of $N$, however, particularly for the model 
with very short range interactions. 

\section{Continuum Simulations}
\label{sec:ContinuumSimulations}
In order to test the predictions of Sec. \ref{sec:LengthDependence}
in more detail, we have also conducted off-lattice Monte Carlo 
simulations of a simple bead-spring model.

Our simulations use a model of flexible polymers with a 
short-range repulsive nonbonded pair potential and a
harmonic bond potential.
The non-bonded pair potential $v_{\rm pair}(r)$ in our 
one-component reference liquid is a purely repulsive shifted 
Lennard-Jones potential, of the form 
\begin{eqnarray}
  v_{\rm pair}(r) & = & \epsilon F(r)
  \nonumber \\
  F(r)            & = & 4 [ (\sigma/r)^{12} - (\sigma/r)^{6}] + 1
  \label{LJfunction}
\end{eqnarray}
for $r$ less than than a cutoff $r_{c} = \sigma 2^{1/6}$, and 
$v_{\rm pair}(r) = F(r)=0$ for $r > r_{c}$. The bond potential 
is a harmonic spring, 
\begin{equation}
   v_{\rm bond}(r) = \frac{1}{2}\kappa(r-l)^{2}
   \quad.
\end{equation}
All of our simulations have used parameter $\epsilon = kT$,
$l = \sigma$, and $\kappa = 400 kT$.

We have simulated chains of of length $N=$ 16, 32, 64, 128, and 
256 at a fixed monomer concentration of $c = 0.7 \sigma^{-3}$.  
For these parameters, we obtain an asymptotic statistical 
segment length $b=1.335$, which was obtained by extrapolating 
the slope of a plot of mean squared end-to-end vector versus 
$N$ to $N=\infty$.  
All simulations use a cubic $L \times L \times L$ 
simulation cell with periodic boundary conditions. We have
simulated systems containing $M=$ 1176, 588 294, 146, and 
146 chains for $N=$16, 32, 64, 128, and 256, respectively. 
This correspond to $L \simeq 29.955 \sigma$ for $N \leq 128$, 
and $L \simeq 37.7$ for $N=256$. 
The Monte Carlo simulations reported here were carried out 
using a combination of hybrid Monte Carlo / Molecular Dynamics 
(MC/MD), reptation, and double-rebridging \cite{dePablo_03} 
moves. 

\begin{figure}[tb]\center
\includegraphics[width=0.45\textwidth,height=!]{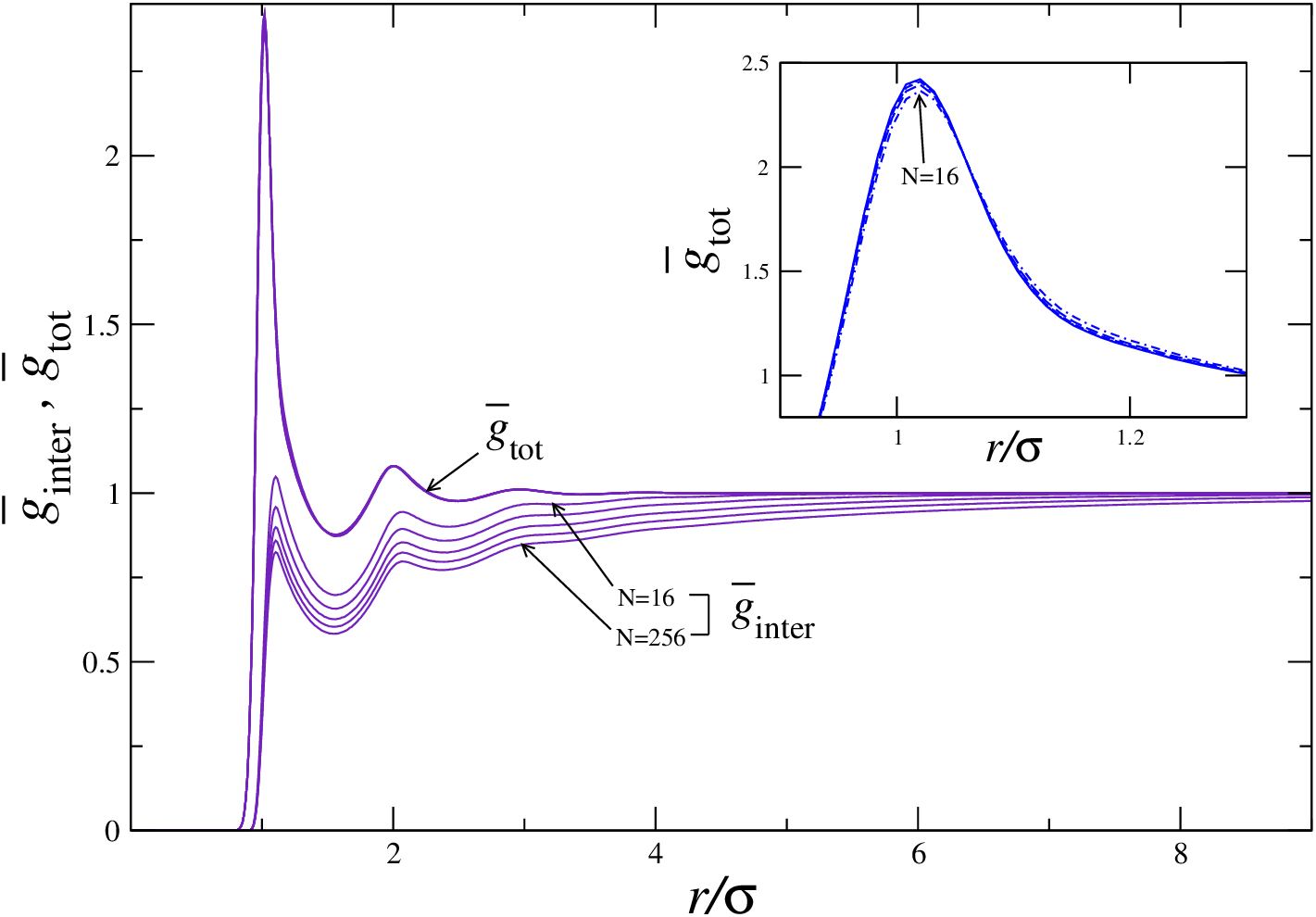}
\caption[Radial Distribution Functions]
{Intermolecular and total radial distribution functions
$\ginterb(\rv)$ and $\gtotb(\rv)$ for chains of length 
$N=$16,\ldots, 256. Results for $\gtotb(\rv)$ for chains 
of different length are indinstinguishable in the main plot. 
Inset: $\gtotb(\rv)$ in an expanded scale, in which the 
slight dependence on $N$ is visible.}
\label{fig:rdf}
\end{figure}

Fig. \ref{fig:rdf} shows the distribution functions $\gtotb(\rv)$ 
and $\ginterb(\rv)$ in the one-component liquid for chains 
of length $N=16,\ldots,256$.  Results for $\gtotb(\rv)$ for 
chains of different length are indistinguishable at the scale 
of the main plot, but the correlation hole in $\ginterb(\rv)$ 
becomes visibly deeper with increasing $N$.  The slight
dependence of $\gtotb(\rv,N)$ on $N$ is visible in the 
inset.

\subsection{Intermolecular Coordination Number}
We consider a perturbation theory for a model of blends in
which the nonbonded pair potential is of the form used by 
Grest and Lacasse \cite{Grest_Lacasse_96}. In this model, 
the pair potential for $i$ and $j$ monomers is taken to 
be a repulsive LJ potential $v_{ij}(r) = \epsilon_{ij}F(r)$, 
where $F(r)$ is given by Eq. (\ref{LJfunction}), in which
$\epsilon_{\As\As} = \epsilon_{\Bs\Bs} = \epsilon$ and 
$\epsilon_{\As\Bs} = \epsilon + \alpha$, 
while the same LJ diameter $\sigma$ and cutoff distance 
$r_{c} = 2^{1/6}\sigma$ is used for all $i$ and $j$. 
For this model,
\begin{equation}
   \ze(s,N) = c\int\!d\rv\; \ginters(\rv,s,N)F(r)
   \quad,
\end{equation}
and $\ze(N)$ is the corresponding average with respect 
to $s$.

\begin{figure}[tb]\center
\includegraphics[width=0.45\textwidth,height=!]{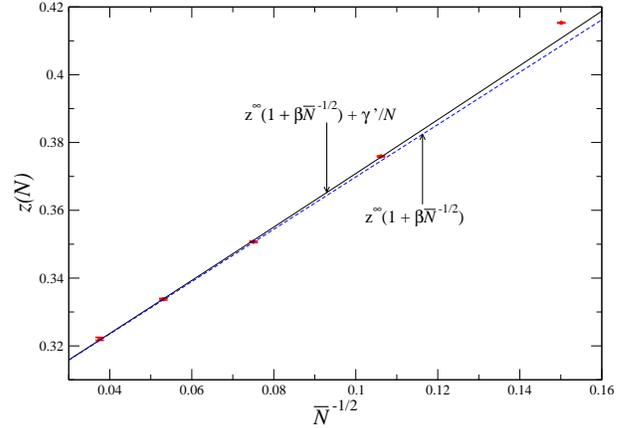}
\caption[$\ze(s,N)$]
{Simulation results for $\ze(N)$ (symbols) vs. $\bar{N}^{-1/2}$,
compared to the prediction of Eq. (\ref{zcAll_predict}), using 
the predicted value of $\beta=(6/\pi)^{3/2}$.  A best fit to
the four longest chain lengths ($N=32,\ldots,256$), shown by 
the solid line, yields parameters $\zeinfty=0.2926$ and 
$\gamma'=0.0361$. The dashed line is the asymptotic line
$\zeinfty[ 1 + \beta \bar{N}^{-1/2}]$, with the same values 
for $\zeinfty$ and $\beta$. Error bars are shown, but are
very small on this scale.}
\label{fig:zcAll}
\end{figure}

Fig. \ref{fig:zcAll} shows our results for $\ze(N)$.
To compare this data to theoretical predictions, we have fit 
values of $\ze(N)$ for all but the shortest chain ($N=16$) to 
Eq.  (\ref{zcAll_predict}). We chose to exclude the data for 
$N=16$ from this fit because the predictions are an asymptotic 
expansion that is expected to be acccurate only for 
sufficiently long chains, and because excluding the shortest
chains from this data substantially improved the quality of 
the fit. The fit for the 4 longest chains, $N=$32,\ldots,256, 
agrees with the data to within our (very small) statistical 
errors.

\begin{figure}[tb]\center
\includegraphics[width=0.45\textwidth,height=!]{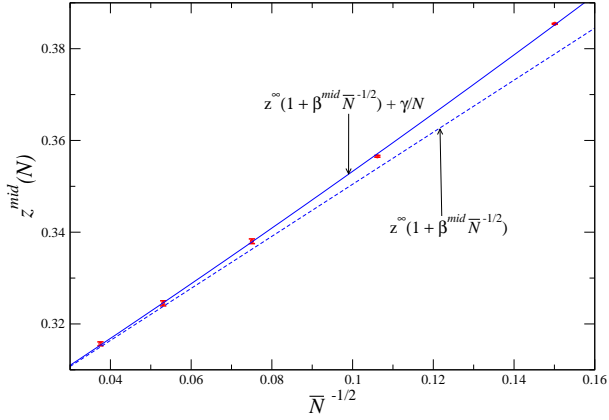}
\caption[$\ze(s,N)$]
{Simulation results for $\ze^{mid}(N)$ (symbols) vs. $\bar{N}^{-1/2}$,
compared to the prediction of Eq. (\ref{zc_mid_fit}), using 
Eq. (\ref{beta_mid_value}) for $\beta^{mid}$.  A best fit, 
shown by the solid line, yields parameters $\zeinfty=0.2937$ 
and $\gamma=0.1014$. The dashed line is the asymptote 
$\zeinfty[ 1 + \beta \bar{N}^{-1/2}]$.}
\label{fig:zc}
\end{figure}

Fig. \ref{fig:zc} shows a corresponding comparison of theoretical 
predictions to simulation results for the quantity
\begin{eqnarray}
    \ze^{mid}(N) \equiv 
    \frac{2}{N}\sum_{s=N/4+1}^{3N/4} \ze(s,N) 
    \quad. \label{zmid_def}
\end{eqnarray}
This is the average of $\ze(s,N)$ over the middle half of each 
chain. This quantity, unlike the average $\ze(N)$ over all monomers, 
excludes contributions from monomers very near the chain ends. By 
using Eq. (\ref{zesN_with_end}) for $\ze(s,N)$, and approximating 
the sum over $s$ in Eq. (\ref{zmid_def}) by an integral over 
$N/4 < s < 3N/4$, we obtain a predicted $N$-dependence
\begin{equation}
   \ze^{mid}(N) = \zeinfty 
   \left [ 1 + \beta^{mid} \bar{N}^{-1/2}
   \right ] + \frac{\gamma}{N} \quad,
   \label{zc_mid_fit}
\end{equation}
in which
\begin{equation}
   \beta^{mid} \equiv (\sqrt{3}-1)\left (\frac{6}{\pi}\right )^{3/2}
   = 1.932
   \quad. \label{beta_mid_value}
\end{equation}
The approximation of a sum over $s$ by an integral gives rise to 
errors of ${\cal O}(N^{-3/2})$, which lie beyond the ${\cal O}(1/N)$ 
accuracy of Eq. (\ref{zesN_with_end}) for the summand $\ze(s,N)$. 
Because the sum over $s$ that defines $\ze^{mid}(N)$ only
includes interior monomers, for which we expect Eq. (\ref{zesN_with_end}) 
for $\ze(s,N)$ to be valid, we may identify the constant $\gamma$ 
in this fit with the constant $\gamma$ in Eq. (\ref{zesN_with_end}).
The prediction fits the data for all $N=16,\ldots,256$ to within
the (small) statistical errors.
A fit of the same data to 
$\zeinfty[ 1 + \beta^{mid}_{fit} \bar{N}^{-1/2}]$, in which 
$\beta^{mid}_{fit}$ is treated as an adjustable parameter, yields 
a slightly worse fit, and a value $\beta^{mid} = 2.1233$ slightly
higher than the predicted value of $\beta^{mid}$.

\begin{figure}[tb]\center
\includegraphics[width=0.45\textwidth=,height=!]{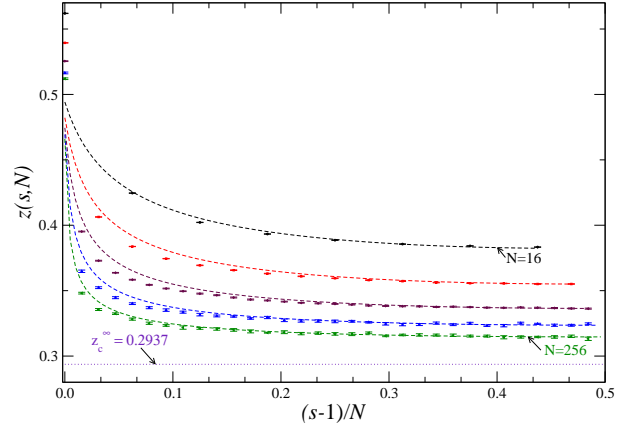}
\caption[$\ze^{mid}(N)$]
{ Simulation results for $\ze(s,N)$ (symbols) vs. $s$, for 
$s=1,\ldots,N/2$ and chains of length $N=$16,\ldots,256, compared 
to the prediction of Eqs. (\ref{zesN_with_end}) and (\ref{zcs_fit}),
shown by dashed lines. Values for the two parameters $\zeinfty$
and $\gamma$ were taken from the fit of $\ze^{mid}(N)$ shown 
in Fig. \ref{fig:zc}. }
\label{fig:pzc}
\end{figure}

Fig. \ref{fig:pzc} shows our simulation results for $\ze(s,N)$ 
for monomers $s=1,\ldots,N/2$ on chains of length $N=$16,\ldots,256.  
Results for each chain length are compared to the predictions of 
Eq. (\ref{zesN_with_end}), shown by dashed lines. Values for the 
two parameters $\zeinfty$ and $\gamma$ have been taken from the 
fit of $\ze^{mid}(N)$ shown in Fig.  \ref{fig:zc}. Because 
Eq. (\ref{domega0sN}) for $\delta \wintras_{\Gauss}(0,s,N)$ was 
derived using continuous Gaussian chain model, and yields a value 
that diverges at both chain ends, we have taken
\begin{equation}
  \delta \wintras_{\Gauss}(0,s,N) =
  \left ( \frac{3}{2\pi} \right )^{3/2} 
  \frac{2}{b^{3}} 
  \left [ \frac{1}{\sqrt{s-a}} + \frac{1}{\sqrt{N+a-s}} \right ]
  \quad, \label{zcs_fit}
\end{equation}
to apply Eq. (\ref{domega0sN}) to a discrete chain with monomer 
indices $s=1,\ldots,N$, with $a=1/2$. Different choices for the 
value of the offset $a$ yield predictions that differ by a
correction of ${\cal O}(N^{-3/2})$. The value $a=1/2$ was chosen
to minimize the discretization error arising from our use of 
integral to approximate a sum over $s$ in the derivation of 
Eq. (\ref{zc_mid_fit}). 

Agreement between this data for $\ze(s,N)$ and 
Eq. (\ref{zesN_with_end}) is excellent for all $N$, and for all 
$s$ except values near the chain ends, for which all analytic 
approximations are expected to fail. 
 
\subsection{Total Coordination Number}
We have also considered a quantity
\begin{equation}
   \ye(s,N) = c\int\!d\rv\; \gtots(\rv,s,N)F(r)
\end{equation}
that depends on the total RDF $\gtots$.
For the blend model considered here, for which $\dV_{AA}(\rv)$, 
the quantity $\ye(N)$ that appears in the perturbation theory
actually vanishes. We have nonetheless considered the quantity 
defined above as a way to test our assumptions about the chain 
length dependence of $\ginters(\rv,s,N)$.

\begin{figure}[tb]\center
\includegraphics[width=0.45\textwidth,height=!]{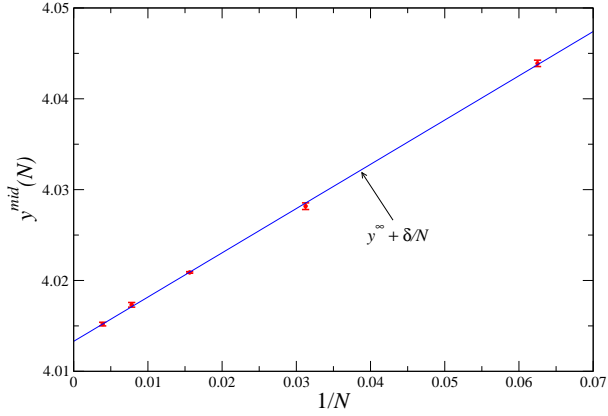}
\caption[$\ye^{mid}(s,N)$]
{Simulation results for $\ye^{mid}(N)$ (symbols) vs. $1/N$. A best 
fit to $\ye^{mid}(N) = \yeinfty + \delta /N$, shown by the solid 
line, yields parameters $\yeinfty = 4.0133$ and $\delta = 0.4872$.}
\label{fig:yc}
\end{figure}

Fig. \ref{fig:yc} shows our results for the 
average
\begin{equation}
    \ye^{mid}(N) \equiv
    \frac{2}{N}\sum_{s=N/4+1}^{3N/4} \ye(s,N) 
\end{equation}
of the total coordination number $\ye(s,N)$ over the middle 
half of each chain, and for $\ye(s,N)$ itself, respectively. 
Results for $\ye^{mid}(N)$ 
have been fit to a predicted form $\ye^{mid}(N) = \yeinfty + \delta/N$,
which follows immediately from Eq. (\ref{yc_predict}) for $\ye(s,N)$
for interior monomers.  Note that the fractional deviations of 
$\ye^{mid}(N)$ from $\yeinfty$ are much smaller than those found 
for $\ze^{mid}(N)$: For the shortest chains, with $N=16$, 
$\ye^{mid}(N)$ deviates from $\yeinfty$ by about 1\%, whereas 
$\ze^{mid}(N)$ deviates from $\zeinfty$ by roughly 30\%.

\begin{figure}[]\center
\includegraphics[width=0.45\textwidth,height=!]{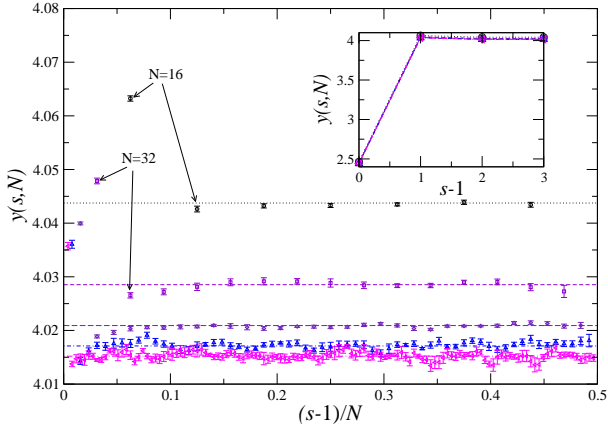}
\caption[$\ye^{mid}(s,N)$]
{Simulation results for $\ye(s,N)$ (symbols) vs. $s$ for
$s=1,\ldots,N/2$ and $N=$16,\ldots,256, compared to the prediction 
$\ye(s,N) = \yeinfty + \delta /N$ (dashed lines). Values for the
parameters $\yeinfty$ and $\delta$ were taken from the fit shown
in Fig. \ref{fig:pyc}.}
\label{fig:pyc}
\end{figure}

In Fig. \ref{fig:pyc}, we compare data for $\ye(s,N)$ for all $s$ 
and $N$ to the functional form predicted in Eq. (\ref{yc_predict}) 
for interior monomers. Here, we have used the values of $\yeinfty$ 
and $\delta$ obtained from the fit shown in Fig. \ref{fig:yc}.
This data clearly confirms that the small deviations $\ye(s,N)$ 
from $\yeinfty$ for interior monomers are independent of $s$ and 
proportional to $1/N$, as expected if the dominant corrections 
to $\yeinfty$ arise from occasional contact of interior monomers 
with end monomers. The inset to Fig. \ref{fig:pyc} shows the much 
larger ${\cal O}(1)$ deviation of $\ye(s,N)$ from $\yeinfty$ for 
the last bead at each chain end.

\section{SCFT and One-Loop Theory}
\label{sec:CoarseGrained}
There are good reasons to believe that a generalized form of SCFT 
becomes increasingly accurate with increasing chain length, and 
is exact in the limit of infinitely long chains. 
The strongest theoretical evidence for this hypothesis comes 
from investigation of corrections to SCFT within the context 
of the renormalized one-loop theory \cite{Morse_Grzywacz_07}, as
discussed below.  The hypothesis that SCFT is exact in the limit 
of long chains is, of course, also consistent with the striking 
success of the Flory-Huggins and (particularly) RPA theories in 
describing experimental data from mixtures of long finite 
polymers. 

The renormalized one-loop theory yields a prediction for the 
free energy per monomer of the form
\begin{equation}
    f = 
    kT \sum_{i}\frac{\volfrac_i}{N_i}\ln \volfrac_{i} 
    + f_{\rm int}(\volfrac,T)
    + f^{*}
   \quad, \label{f_one_loop_form}
\end{equation}
in which $f^{*}$ is a one-loop correction to generalized 
Flory-Huggins theory. For symmetric models of the type
considered here, the predicted correction $f^{*}$ is a
function of the form
\begin{equation}
   f^{*}(\volfrac,\chie,N) = 
   \frac{kT}{N\bar{N}^{1/2}} 
   \hat{f}^{*}(\volfrac, \chie N) 
   \label{f_star_nondim}
\end{equation}
where $\hat{f}^{*}$ is a dimensionless function of $\volfrac$
and $\chie N$. Here, $\chie$ is an effective interaction 
parameter that is related to $\Delta f_{\rm int}(\phi,T)$ by
Eq. (\ref{chie_def}).  The SCF interaction free energy $f_{\rm int}$, 
which is required as an input to the theory, can have an arbitrary 
composition dependence.  To compare one-loop predictions to our 
perturbation theory of symmetric mixtures, we must also allow 
$f_{\rm int}$ to depend upon the parameter $\alpha$ of the 
underlying microscopic model.

\subsection{SCFT}
Because the one-loop contribution $f^{*}$ decreases as $N^{-1/2}$ 
with increasing chain length, the SCF contribution $f_{\rm int}$ 
may be identified by considering the limit $N \rightarrow \infty$ 
of the true free energy $f$. As discussed in the introduction, 
an expansion of $f_{\rm int}$ to first order in $\alpha$ may thus 
be obtained by simply taking the limit $N \rightarrow \infty$ of 
the corresponding first order expansion of $f$, which is given 
in Eq.(\ref{DeltaFPerturb}). This yields 
\begin{equation}
  f_{\rm int} = f_{0} + 
  \frac{1}{2}\alpha \yeinfty + 
  \alpha \zeinfty \volfrac_{A}\volfrac_{B}
  \quad.
  \label{Fint}
\end{equation}
An expression for the corresponding contribution 
$\Delta f_{int}$ to the free energy of mixing is given in
Eq. (\ref{DeltaFint}). This result yields an expansion of
effective SCF interaction parameter $\chie$, to first 
order in $\alpha$, as 
\begin{equation}
    \chie = \frac{\alpha\zeinfty}{kT}
    \quad. \label{chie_zeinfty}
\end{equation}
The fact that this expansion for $\chie$ is independent of 
composition is a special feature of the expansion of this
class of symmetric models to first order in $\alpha$: We do
not expect it to survive any generalization to structurally 
asymmetric models or to higher order in $\alpha$.  Nothing 
rigorous can be said about the temperature dependence, 
even in this first order expansion, because the expansion 
with respect to $\alpha$ has been carried out at constant 
$T$, and $\zeinfty$ thus has an unknown dependence on $T$. 
The coefficient $\zeinfty$ would be independent of $T$ 
only in an athermal reference system, such as a lattice 
model with no vacancies or a model of tangent hard spheres. 

\subsection{One-Loop Contribution}
Both the one-loop theory for $f$ and the simple perturbative
expansion of $f$ predict corrections to the SCF free energy,
as defined above, that are of order $\bar{N}^{-1/2}$. The 
one-loop theory is simply not a perturbation theory, since 
it predicts a correction $f^{*}(\phi,\chie N)$ that is a 
nonlinear function of $\chie N$, and that exhibits singular 
behavior near the spinodal.  We can test whether one-loop 
predictions are {\it consistent} with our analysis of a 
more microscopic perturbation theory, however, by 
considering the predictions of the one-loop theory for the 
derivative $\theta = \partial f/\partial \alpha$ at $\alpha=0$, 
and comparing expressions for the ${\cal O}(\bar{N}^{-1/2})$
contribution to this coefficient. 

The one-loop expression for $\theta \equiv \partial f/\partial \alpha$ 
at $\alpha=0$ can be expressed as a sum 
\begin{equation}
   \frac{\partial f}{\partial \alpha} = 
   \frac{\partial f_{int}}{\partial \alpha} +
   \frac{\partial f^{*}}{\partial \alpha} 
   \label{theta_FH_theta_star}
   \quad.
\end{equation}
The SCF contribution is simply
\begin{equation}
   \left . \frac{\partial f_{int}}{\partial \alpha} \right |_{\alpha=0}
   = \frac{1}{2}\yeinfty + \zeinfty \volfrac_{\As}\volfrac_{\Bs}
   \label{theta_FH}
\end{equation}
and $\theta^{*} \equiv \partial f^{*}/\partial \alpha$. 
The one-loop correction is of the form
\begin{equation}
   \frac{\partial f^{*}}{\partial \alpha} = 
   \frac{\partial \chie(\volfrac, \alpha)}{\partial \alpha} 
   \frac{\partial f^{*}(\volfrac, \chie, N)}{\partial \chie } 
   \quad.
   \label{theta_star_formal} 
\end{equation}
It is straightforward to show, by using the functional form given
in Eq. (\ref{f_star_nondim}), that this correction is proportional 
to $\bar{N}^{-1/2}$. 

In the accompanying paper, we explicitly calculate the required
derivative of the one-loop correction $f^{*}$, and find that
\begin{equation}
   \left . \frac{\partial f^{*}}{\partial \alpha} \right |_{\alpha=0} = 
   v \delta\wintrab_{\Gauss}(\rv=0,N) \volfrac_{\As}\volfrac_{\Bs} 
   \quad.
\end{equation}
Combining this with Eqs. 
(\ref{theta_FH_theta_star}-\ref{theta_star_formal}) and 
Eq. (\ref{chie_zeinfty}) for $\chie$ yields
\begin{equation}
   \left . \frac{\partial f}{\partial \alpha} \right |_{\alpha=0} 
   = \frac{1}{2}\yeinfty
   + \zeinfty \volfrac_{\As}\volfrac_{\Bs}
   \left [ 1 + v\delta\wintrab_{\Gauss}(\rv=0,N) \right ]
   \quad.
\end{equation}
This is identical to the expression obtained in Sec. 
\ref{sec:LengthDependence}.  

We thus conclude that the one-loop theory implicitly contains 
a correct description of the $N$-dependence of the correlation 
hole. This is enough to guarantee that the one-loop theory will
yield a very accurate description of corrections to SCFT in
weakly non-ideal symmetric blends, with $\chie N \ll 1$. In
an accompanying paper, we confirm that this is true by comparing 
one-loop predictions to simulation results for composition 
fluctuations in such blends. There, we also examine the accuracy 
of the one loop theory for larger values of $\alpha$, up to the
critical value, for which perturbation theory becomes inadequate.

\section{Conclusions}
A simple physical picture has been given for how intra- and 
inter-molecular correlation functions vary with chain length in 
a polymer melt. A theory based on this picture is in excellent
agreement with computer simulation results.  The structure of 
a one-component melt is related by perturbation theory to the 
free energy of mixing in corresponding structurally symmetric 
blends. The ${\cal O}(N^{-1/2})$ contribution to the depth of 
the intramolecular correlation hole in the melt of finite 
chains leads to a slightly higher free energy of mixing in 
mixtures of shorter chains. This simply reflects the fact that 
monomers on shorter chains are less strongly screened from
contact with other chains. Perturbation theory may be used
to estimate the SCF interaction free energy appropriate for 
comparison of SCF theory to simulations, by identifying SCF
theory with the $N \rightarrow \infty$ limit of the perturbation 
theory. If this prescription is used identify SCF parameters, 
the predictions of the one-loop theory for corrections to
SCF theory is found to be consistent with the perturbation
theory presented here, insofar as both theories give identical
results for a ${\cal O}(N^{-1/2})$ correction to the apparent 
interaction parameter in weakly non-ideal symmetric mixtures, 
with $\chi N \ll 1$. 

\acknowledgments{
The simulations reported here were completed with the 
resources of the Minnesota Supercomputer Institute.
}

\end{document}